\documentclass[twocolumn, iop, apj, twocolappendix, numberedappendix, appendixfloats]{emulateapj}
\usepackage{CJKutf8}
\usepackage{threeparttable}
\usepackage{multirow}
\usepackage{times}
\usepackage{enumitem}
\usepackage{url} 
\usepackage{color}
\usepackage{amsmath,bm}
\usepackage{subfigure}
\usepackage{natbib}
\usepackage{graphicx}
\usepackage{graphics}
\usepackage{hyperref}                             
\usepackage{amsmath}
\usepackage{bm}
\usepackage{subfigure}
\usepackage{array}
\usepackage{booktabs}
\usepackage{lineno}

\shorttitle{The tilt of the velocity ellipsoid}

\shortauthors{The tilt of the velocity ellipsoid}

\begin{document}

\title{The Tilt of the Velocity Ellipsoid in the Milky Way Disk with GALAH-Gaia Data}

\author{Zhenzhen He\textsuperscript{1}}

\author{Ruilan Yang\textsuperscript{3, 4}}

\author{Hui Li\textsuperscript{2,4}}

\author{Weixiang Sun\textsuperscript{2,4}}

\altaffiltext{1}{Henan Police College, Zhengzhou 450046, People’s Republic of China}

\altaffiltext{2}{Department of Astronomy, Tsinghua University, Beijing 100084, People’s Republic of China; {\it sunweixiang@tsinghua.edu.cn {\rm (WXS)}}; {\it hliastro@tsinghua.edu.cn {\rm (HL)}}}

\altaffiltext{3}{Henan Police College, Zhengzhou 450046, People’s Republic of China; {\it yang.rl.mail@gmail.com {\rm (RLY)}}}

\altaffiltext{4}{Corresponding authors}

\begin{abstract}

We investigate the metallicity dependence of the tilt of the velocity ellipsoid using more than 572,000 stars selected from GALAH and Gaia.
We find that the tilt angles of disk stars can be well described by the relation $\alpha = \alpha_{0}\,\mathrm{arctan}(Z/R)$, and the $\alpha_{0}$ differ significantly among different disk populations, with $\alpha_{0}$ are 0.78 $\pm$ 0.05, 0.86 $\pm$ 0.07, and 0.77 $\pm$ 0.07 for the whole sample, the Low-[$\alpha$/Fe] and the High-[$\alpha$/Fe] populations, respectively.
These results indicate that Galactic disk populations are globally tilted toward cylindrical alignment, implying that the gravitational potential of the Galactic baryonic disk plays a dominant role in shaping the kinematics of dynamically relaxed stars.
Our detailed analysis further reveals a metallicity dependence of $\alpha_{0}$ for various populations, with the Low-[$\alpha$/Fe] and the High-[$\alpha$/Fe] populations are respectively $\alpha_{0}$ = 0.12\,$\mathrm{sin}$\,(3.72\,[Fe/H]\,+\,1.09)\,+\,0.75 and $\alpha_{0}$ = 0.20\,$\mathrm{sin}$\,(2.99\,[Fe/H]\,+\,1.90)\,+\,0.71.
These results indicate that the metallicity dependence observed in the Low-[$\alpha$/Fe] population may be primarily associated with resonances induced by the Galactic bar, while the behavior of the High-[$\alpha$/Fe] population may also have been affected by additional dynamical and evolutionary processes, such as violent heating associated with mergers, accretion, and/or the formation of these stars in an environment shaped by mergers of gas-rich systems and/or a turbulent ISM.

\end{abstract}


\section{Introduction}

Constraining the mass distribution of the Milky Way is crucial for understanding its formation history and dynamical evolution \citep[e.g.,][]{Binney2008, Smith2009, Everall2019}.
While direct observations can probe the luminous component of the Galaxy, this constitutes only a small fraction of the total mass \citep[e.g.,][]{Huang2016, Eilers2019, Ablimit2020}.
As a result, indirect approaches are essential, among which stellar kinematics provides a particularly sensitive tracer of the underlying gravitational potential \citep[e.g.,][]{Eddington1915, Smith2009, Hagen2019}.

The velocity ellipsoid of stars contains key information about the gravitational field.
Its orientation, commonly expressed through the tilt angle, can be described as $\alpha$ = $\alpha_{0}$\,$\mathrm{arctan}$\,($Z$/$R$), indicates whether stellar motions are closer to spherical or cylindrical alignment \citep[e.g.,][]{Binney2014, Budenbender2015}. Spherical alignment corresponds to $\alpha_0 = 1.0$, whereas $\alpha_0 < 1.0$ implies a tilt toward the cylindrical plane, likely influenced by the baryonic disk potential \citep[e.g.,][]{Hagen2019, Everall2019}. Observational studies have demonstrated that $\alpha_0$ varies systematically with location and stellar population: local RAVE stars show intermediate alignment \citep{Binney2014}, SEGUE G-dwarfs tend toward more spherical orientation \citep{Budenbender2015}, and stars at higher Galactic altitudes generally approach spherical alignment, revealing a dependence on Galactic height \citep{Everall2019}. Distinctions between disk components are also evident, with Low-[$\alpha$/Fe] and High-[$\alpha$/Fe] populations exhibiting different $\alpha_0$ values, reflecting differences in their spatial distributions and kinematics. \citep[e.g.,][]{Binney2014, Hagen2019, Sun2023}.

Beyond spatial variations, recent studies suggest that $\alpha_0$ may oscillate with Galactocentric radius, potentially tracing dynamical resonances associated with the Galactic bar \citep{Sun2023}.
These findings indicate that the velocity ellipsoid encodes not only the overall gravitational potential but also the present-day non-axisymmetric perturbations of the Galactic disk.
However, previous work has often been limited either by the size of the samples or by insufficient precision in stellar metallicity measurements, restricting the ability to characterize the velocity ellipsoid across chemically distinct populations \citep[e.g.,][]{Hagen2019, Everall2019, Sun2023}.

The GALAH survey enables a systematic investigation of the velocity ellipsoid across stellar populations using high-precision stellar chemical abundances \citep[e.g.,][]{De Silva2015}. Combining kinematics with chemistry allows us to examine how these populations respond to the Galactic potential, providing new insights into the formation and evolution of the Milky Way disk.

In this study, we use GALAH data to perform a detailed analysis of the velocity ellipsoid across the Galactic disk, focusing on its dependence on stellar metallicity and population, in order to improve the constraint on the structural and dynamical properties of the Milky Way.

The paper is organized as follows. In Section 2, we introduce the data used in this paper, and present the results and discussions in Section 3. Finally, we summarize our conclusions in Section 4.

\section{Data}

In this work, we adopt the standard Galactocentric cylindrical coordinate system ($R$, $\phi$, $Z$), where the associated velocity components are represented by $V_{R}$, $V_{\phi}$, and $V_{z}$, respectively.
The three-dimensional positions and velocities of stars are calculated by adopting a Solar Galactocentric distance of $R_{\odot}$ = 8.34\,kpc \citep{Reid2014}, Solar motions of ($U_{\odot}$, $V_{\odot}$, $W_{\odot}$),$=$,$(13.00, 12.24, 7.24)$ km\,s$^{-1}$ \citep{Schonrich2018}, together with a local circular velocity of $V_{c,0}$ = 238.0 km\,s$^{-1}$ \citep[e.g.,][]{Reid2004, Schonrich2010, Schonrich2012, Reid2014, Huang2015, Huang2016, Bland-Hawthorn2016}.

The sample with over 572,000 stars from GALAH survey \citep[e.g.,][]{De Silva2015, Buder2021} is used in this paper, with the uncertainties of the line-of-sight velocity ($V_{\rm r}$), log\,$g$, $T_{\rm eff}$,  [Fe/H] and [$\alpha$/Fe], are typically around, 0.34\,km\,s$^{-1}$, 0.07\,dex, 50\,K, 0.06\,dex and 0.05\,dex \citep[e.g.,][]{Buder2021}.
The stellar distances are estimated from Gaia DR3 parallaxes through a Bayesian technique similar to that described by Bailer-Jones ({\color{blue}{2015}}).
The typical uncertainty of the derived distances is around 5\%--10\%.

The astrometric information for the sample stars is obtained from Gaia DR3 \citep[e.g.,][]{Gaia2022a, Gaia2022b, Recio2022}.
To improve the robustness of the kinematic analysis and minimize the impact of distance uncertainties on the derived velocities, we further apply quality constraints requiring spectral signal-to-noise ratios (SNR) $>$ 20 and relative distance uncertainties (defined as the ratio of the distance uncertainty to the distance) to satisfy $\Delta s$ $<$ 10\%.
To further assess the sensitivity of our results to distance uncertainties, we repeated the analysis using different $\Delta s$ thresholds and by propagating the distance uncertainties through Monte Carlo perturbations.
The corresponding results are summarized in Table A1.

\begin{figure}[t]
\begin{center}
\includegraphics[width=8.5cm]{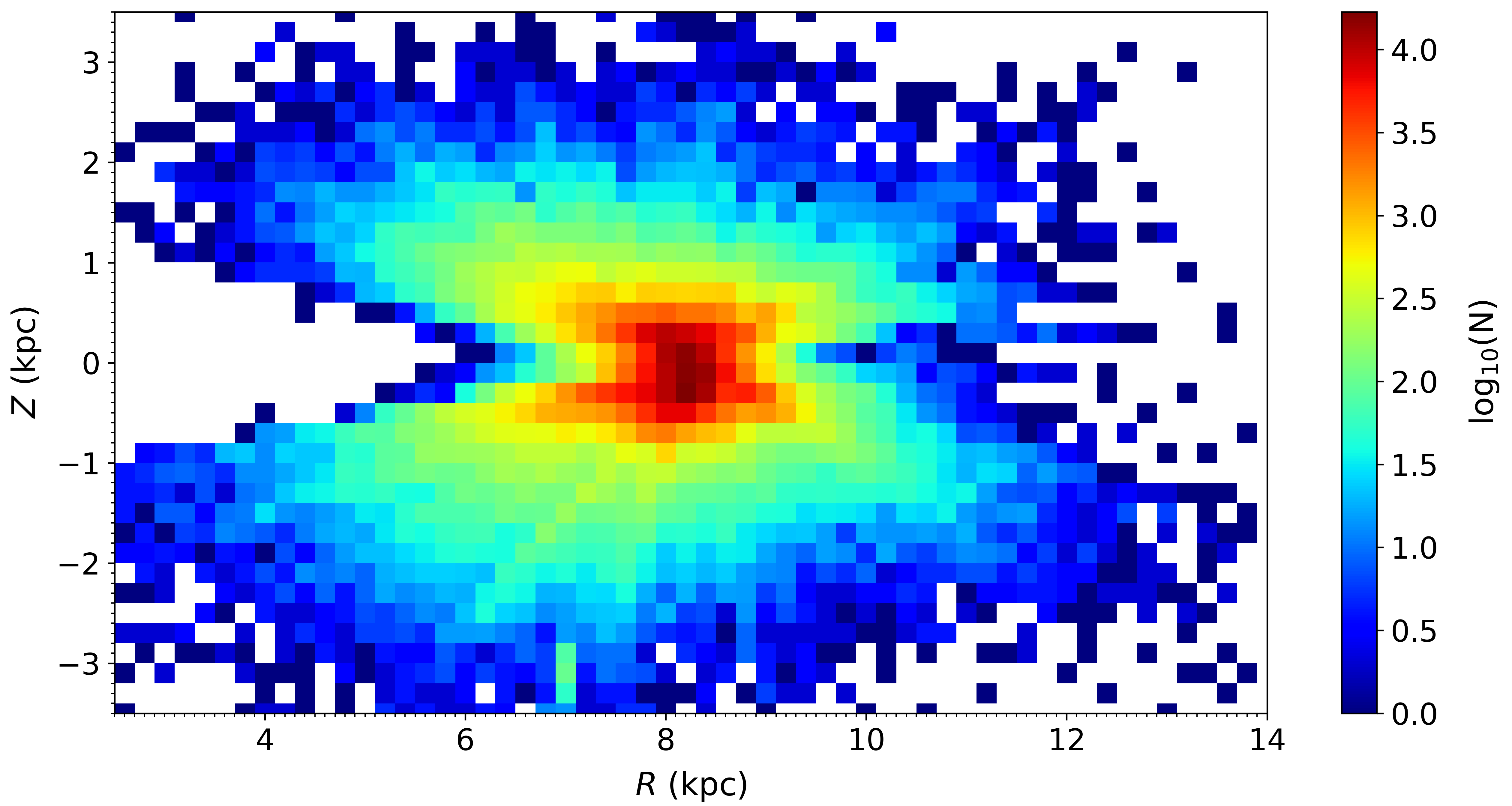}
\caption{The spatial distribution of the sample stars in the $R$--$Z$ phase space, binned in 0.2\,kpc intervals along both axes and colour-coded by stellar density.}
\end{center}
\end{figure}

The velocity dispersions $\sigma$ within each spatial bin are derived after applying a 3$\sigma$-clipping procedure to remove possible outliers.
To reduce contamination from halo populations, we additionally require stellar metallicities of [Fe/H]\,$\geq$\,$-$1.0\,dex \citep[e.g.,][]{Hayden2020, Sun2020}.
After applying all selection criteria, more than 362,000 stars remain in the final sample.
The spatial distribution of the selected stars is presented in Fig.\,1.

\begin{figure*}[t]
\centering

\subfigure{
\includegraphics[width=17cm]{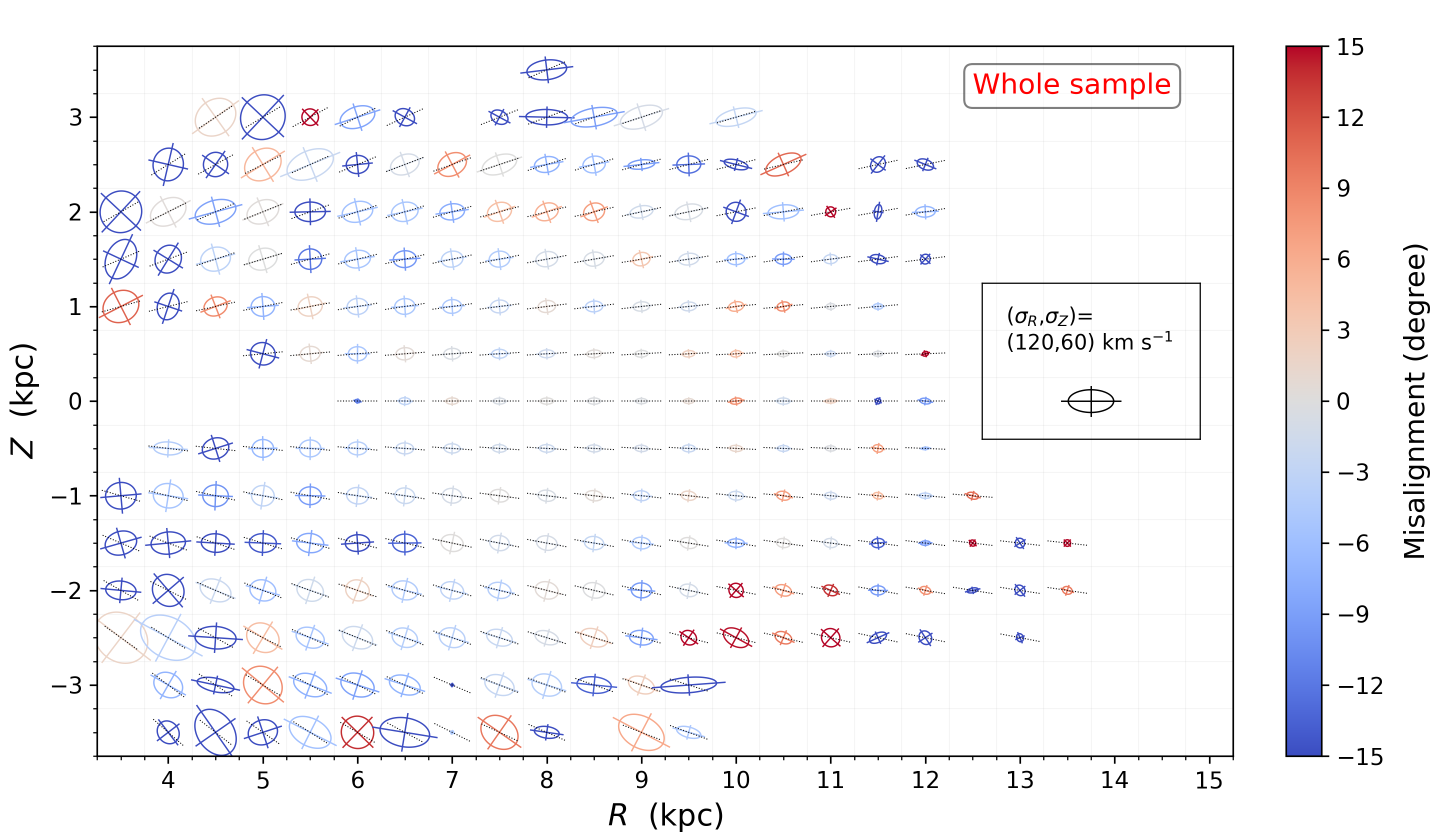}
}

\caption{The distribution of the velocity ellipsoid in the $R$--$Z$ phase space for the whole GALH-Gaia sample, with a bin size of 0.5\,kpc along both axes, and at least 5 stars in each bin.
The velocity ellipses are colour-coded according to their misalignment from spherical alignment.
For comparison, the dotted grey line drawn through each ellipse indicates the orientation expected for spherical alignment.
The size of each ellipse scales with the corresponding velocity dispersion in each spatial bin.
The black rectangle inset illustrates a reference velocity ellipsoid with non-tilted distribution and ($\sigma_{R}$,\,$\sigma_{Z}$)=(120,\,60)\,km\,s$^{-1}$.}
\end{figure*}

\begin{figure*}[t]
\centering

\subfigure{
\includegraphics[width=17cm]{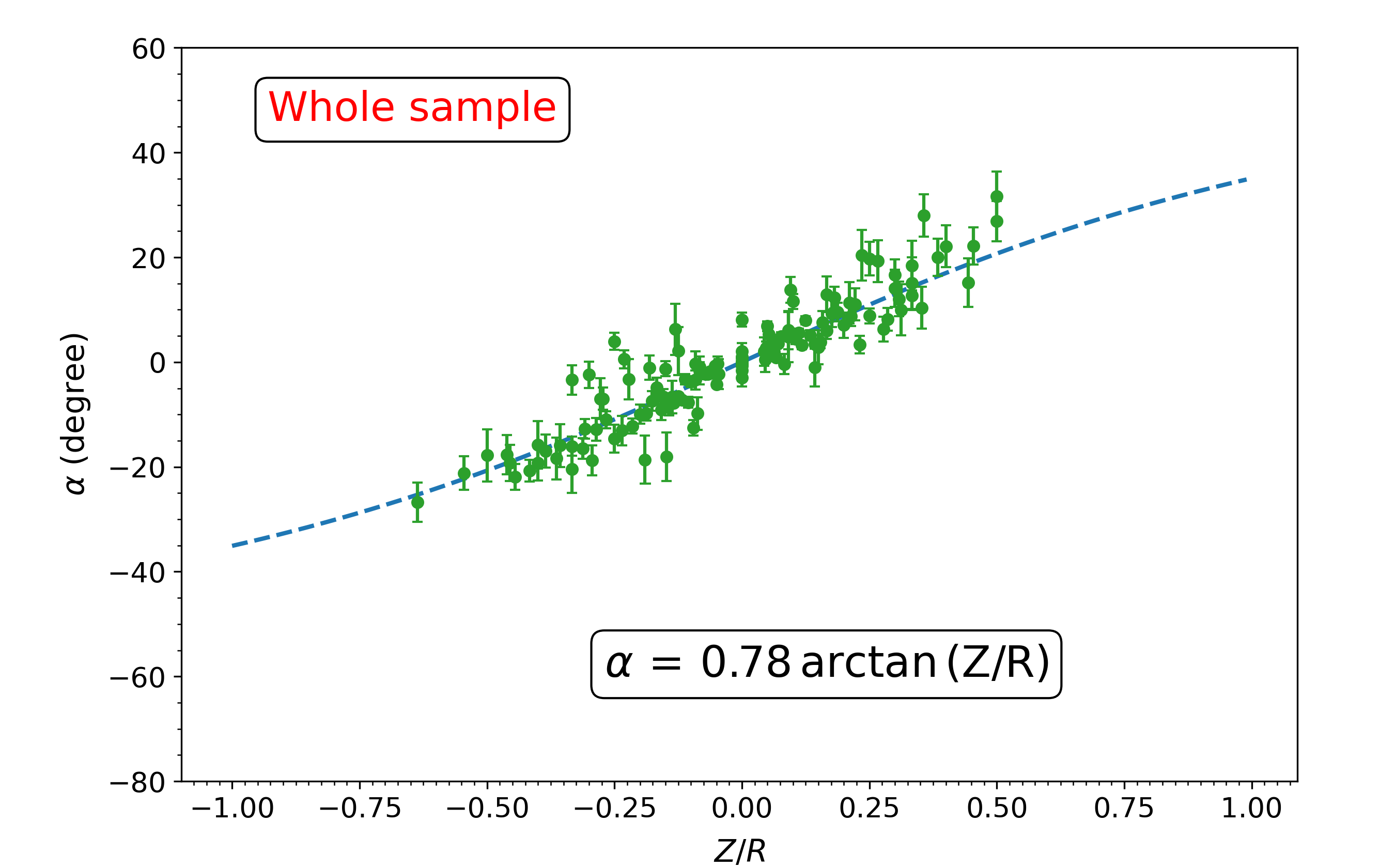}
}

\caption{The tilt angle $\alpha$ as a function of $Z$/$R$, and the red dotted line represents the best fit using Equation (3).}
\end{figure*}

\begin{figure*}[t]
\centering

\subfigure{
\includegraphics[width=17.5cm]{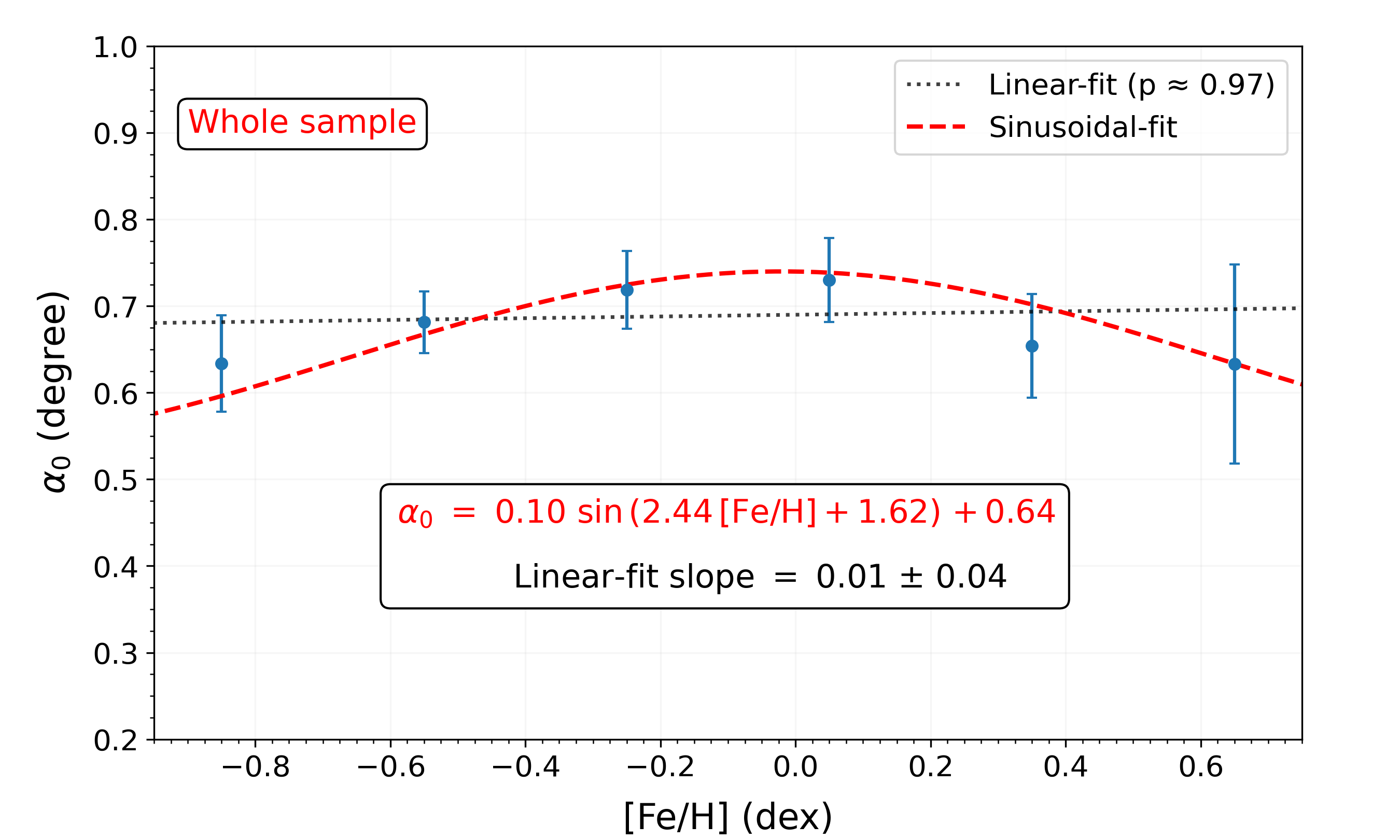}
}

\caption{$\alpha_{0}$ as a function of [Fe/H] for the whole sample stars.
The black dotted line represents a linear fit, yielding a slope of around 0.01 $\pm$ 0.04, which is consistent with zero.
The red dashed line shows a phenomenological sinusoidal fit with Equation (4).}
\end{figure*}

\begin{figure*}[t]
\centering

\subfigure{
\includegraphics[width=17cm]{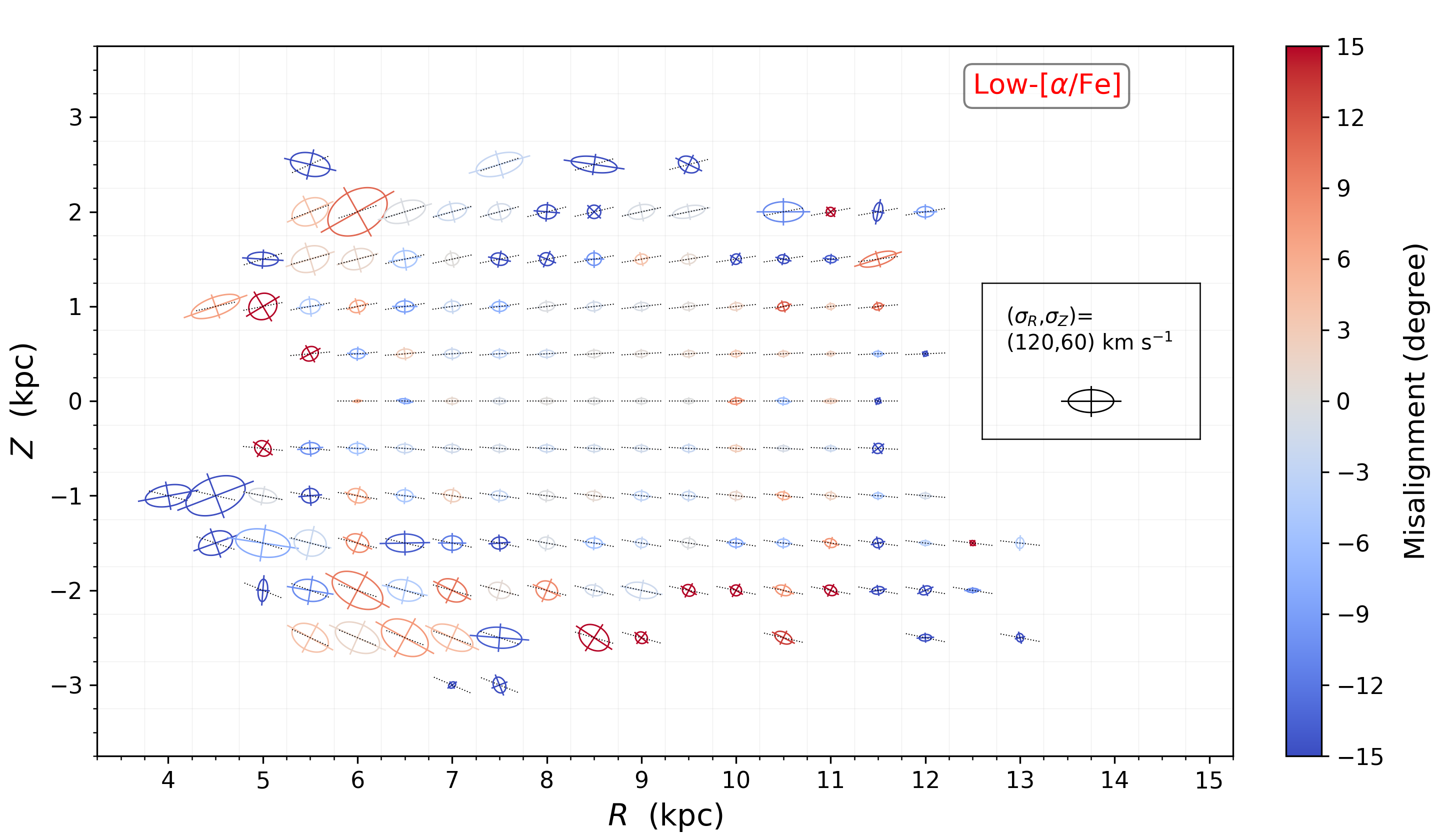}
}

\subfigure{
\includegraphics[width=17cm]{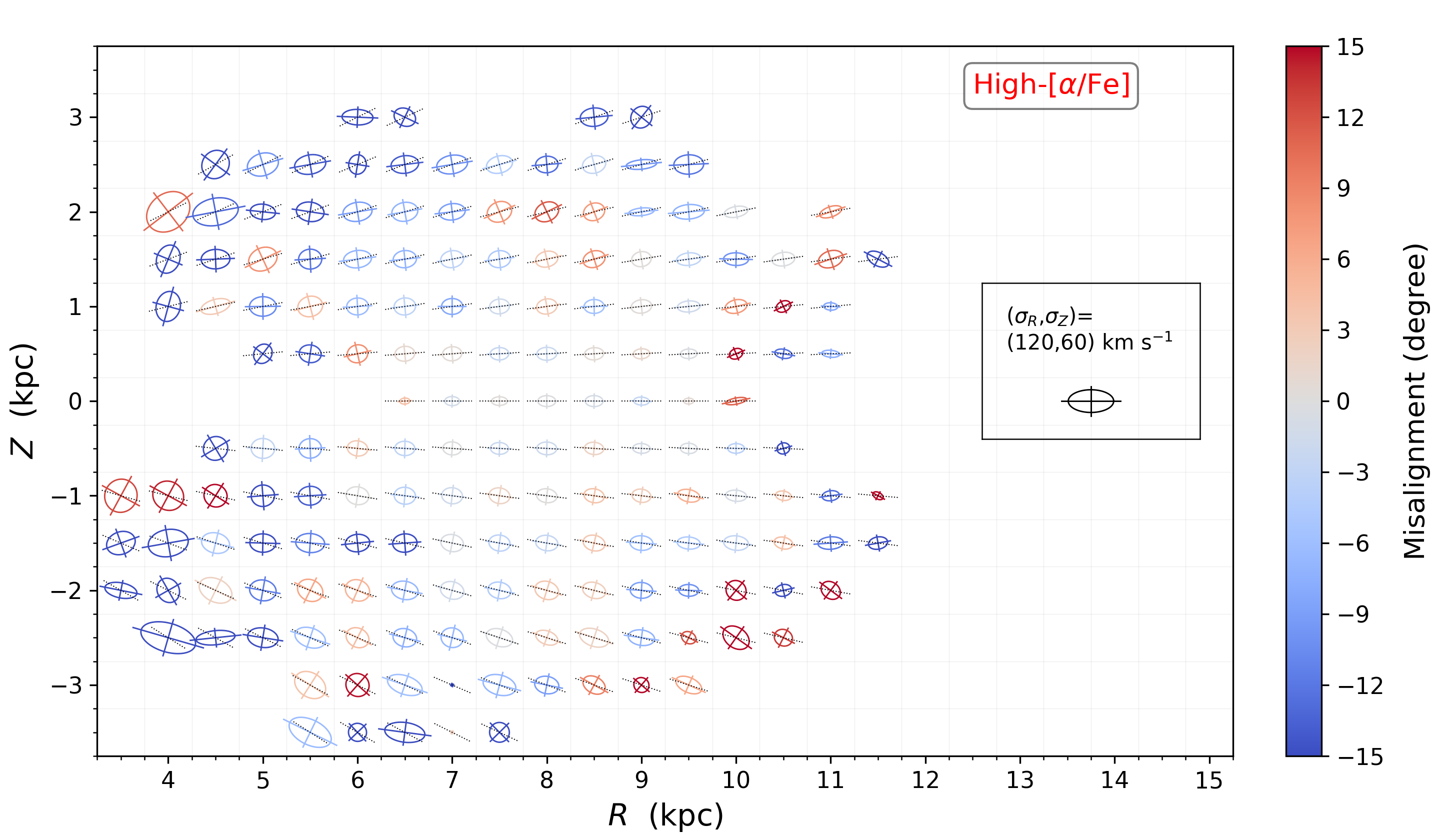}
}

\caption{The distribution of the velocity ellipsoid in the $R$--$Z$ phase space for Low-[$\alpha$/Fe] (upper panel) and High-[$\alpha$/Fe] (bottom panel) populations.}
\end{figure*}

\begin{figure*}[t]
\centering

\subfigure{
\includegraphics[width=17.5cm]{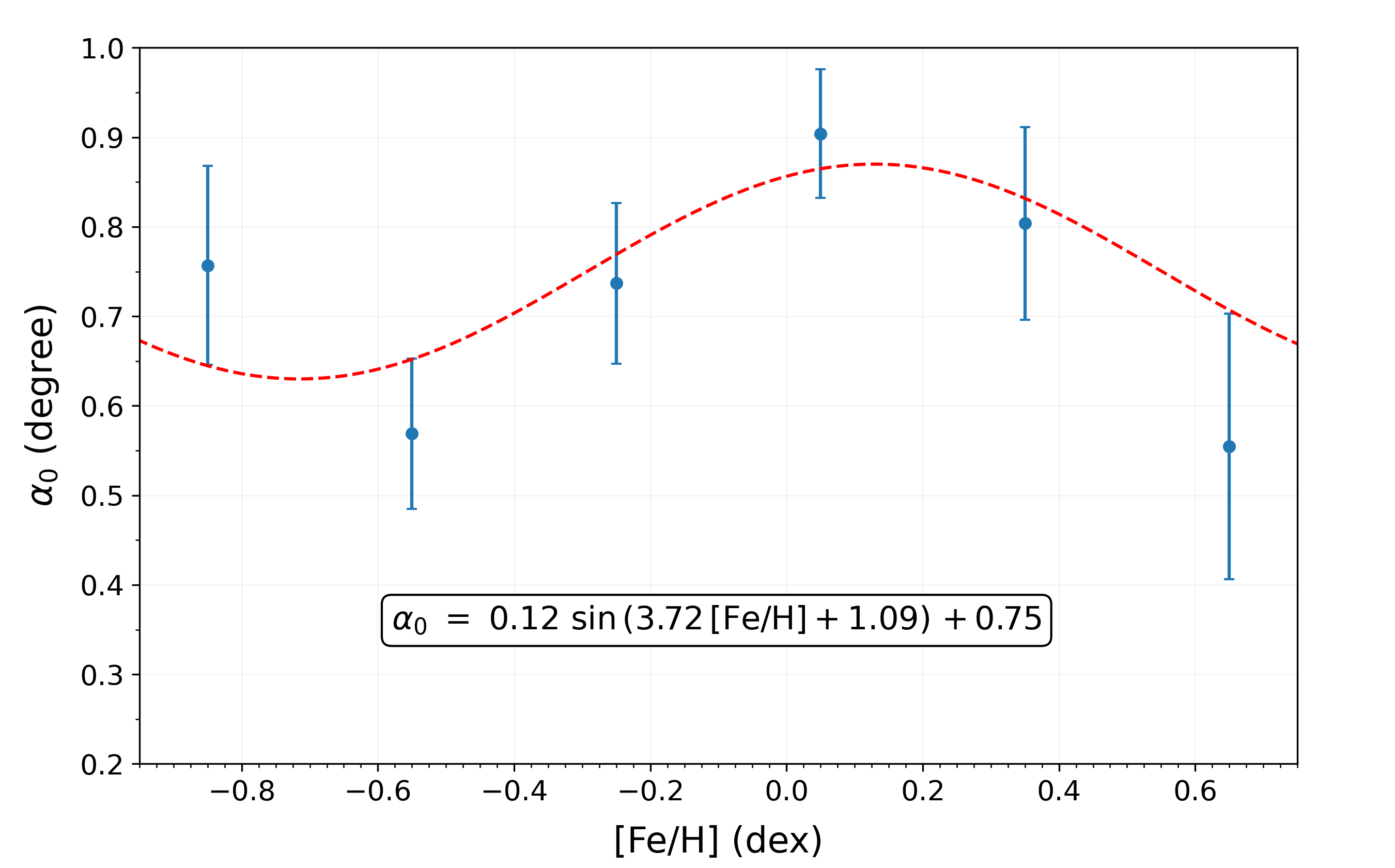}
}

\subfigure{
\includegraphics[width=17.5cm]{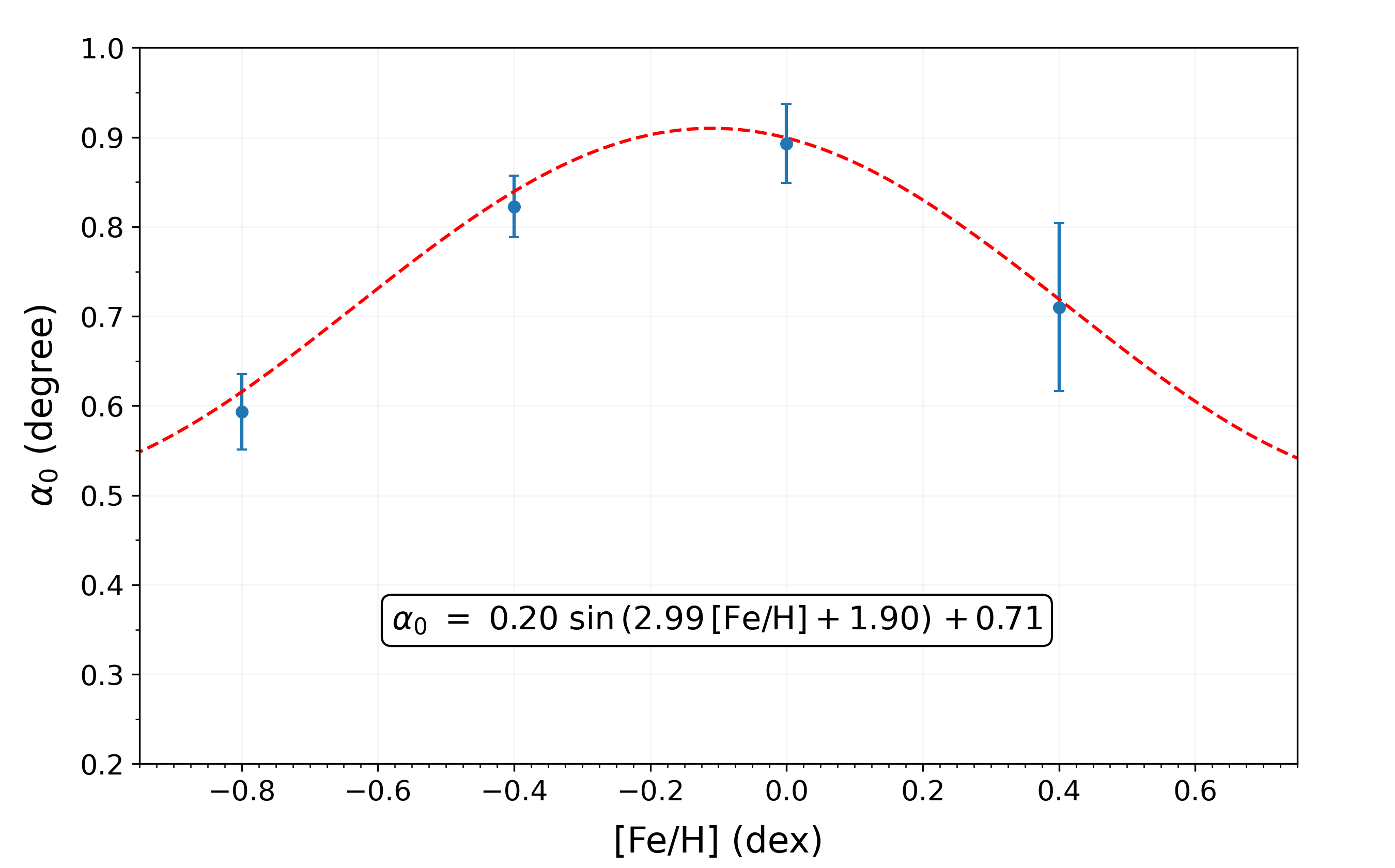}
}

\caption{$\alpha_{0}$ as a function of [Fe/H] for Low-[$\alpha$/Fe] (upper panel) and High-[$\alpha$/Fe] (bottom panel) populations. 
The red dotted line in each panel represents the best fit with Equation (4).}
\end{figure*}

\begin{table*}

\caption{The properties of thin/thin disk, mono-age and mono-[$\alpha$/Fe]-[Fe/H] populations.}

\centering
\setlength{\tabcolsep}{3mm}{
\resizebox{2.\columnwidth}{!}{
\begin{tabular}{llllllllll}
\hline
\hline
\specialrule{0em}{7pt}{0pt}
Name                                         &           $\langle$ [$\alpha$/Fe] $\rangle$  &        $\langle$ [Fe/H] $\rangle$     &     $\langle$ R $\rangle$      &      $\sigma_{R}$            &     $\sigma_{Z}$    &  $\alpha_{0}$  & Number     &    n\,(ratio) \\[0.07cm]
                                             &                (dex)               &        (dex)       &     (kpc)     &        (km\,s$^{-1}$)           &    (km\,s$^{-1}$)       &  (degree)  &     &               \\
\specialrule{0em}{7pt}{0pt}
\hline
\specialrule{0em}{7pt}{0pt}
Low-[$\alpha$/Fe]                                    &                 0.020                  &         $-$0.11          &     8.18       &      37.28               &          19.14       &  0.86 $\pm$ 0.07 &  233,244 &       64.42\%  \\[0.3cm]
High-[$\alpha$/Fe]                                  &                 0.23                  &         $-$0.37        &     7.62       &      55.59               &          37.19       &  0.77 $\pm$ 0.07  &  67,327  &       18.59\%  \\
\specialrule{0em}{7pt}{0pt}
\hline 
\specialrule{0em}{7pt}{0pt}
\end{tabular}}
}
\label{tab:datasets}
\end{table*}

\section{Result and Discussion}

The orientation of the velocity ellipsoid is characterized through the tilt angle between two orthogonal velocity components, following methods widely adopted in previous Galactic dynamical studies \citep[e.g.,][]{Smith2009, Sun2023}.
For a pair of velocity components $v_{i}$, $v_{j}$, the corresponding tilt angle $\alpha_{ij}$ is determined from the covariance tensor according to:

\begin{equation}
\label{eq:tiltangle1}
    \tan(2 \alpha_{ij}) = \frac{2 \mathrm{cov}(v_i, v_j)}{\mathrm{var}(v_i) - \mathrm{var}(v_j)} = \frac{2\sigma_{ij}^{2}}{\sigma_{ii}^{2} - \sigma_{jj}^{2}}
\end{equation}
where

\begin{equation}
\mathrm{cov}(v_i, v_j) \equiv \sigma^2_{ij} \equiv
\left\langle
(v_i - \langle v_i \rangle)(v_j - \langle v_j \rangle)
\right\rangle
\label{eq:cov}
\end{equation}
Here, the angle brackets denote averages over the stellar phase-space distribution function \citep[e.g.,][]{Binney2008, Smith2009, Evans2016, Sun2023}.

The tilt angle $\alpha_{ij}$ describes the orientation of the projected velocity ellipsoid on the $ij$ plane and is measured from the $i$-axis toward the major axis of the ellipse \citep[e.g.,][]{Binney1998}.
By definition, positive values correspond to counter-clockwise rotation, yielding an allowed range between $-$45$^{\circ}$ to $+$45$^{\circ}$.
In the Galactocentric cylindrical coordinate system($R$, $\phi$, $Z$), a tilt angle of $\alpha$ = 0$^{\circ}$ corresponds to a velocity ellipsoid whose major (direction of radial velocity, $V_{R}$) and minor (direction of vertical velocity, $V_{Z}$) axis are respectively fully aligned with $R$-axis and $Z$-axis.

To investigate the dependence of the velocity ellipsoid tilt on stellar populations,  the sample is further divided into low-[$\alpha$/Fe] and high-[$\alpha$/Fe] subsamples following previous chemically motivated classifications \citep[e.g.,][]{Bensby2014, Lee2011, Brook2012, Haywood2013, Recio-Blanco2014, Guiglion2015}.
After applying all selection criteria, 104,935 and 18,566 stars are retained in the low-[$\alpha$/Fe] and high-[$\alpha$/Fe] populations, respectively.
The basic properties of the two populations are summarized in Table\,1.

\subsection{The Velocity Ellipsoid distribution and its tilt for the whole sample}

The velocity ellipsoid properties of the whole sample stars are presented in Fig.\,2--4.
Fig.\,2 displays the distribution of the velocity ellipsoid of the whole sample stars, with color-coded by the degree of misalignment relative to spherical alignment.
The misalignment is defined as $\alpha_{sph}$ $-$ $\alpha$ for stars at $Z$ $<$ 0.0\,kpc, and $\alpha$ $-$ $\alpha_{sph}$ for stars at $Z$ $\geq$ 0.0\,kpc.
Here $\alpha_{sph}$ denotes the tilt angle expected for fully spherical alignment.
In the figure, blue and red colors correspond to velocity ellipsoids that are relatively more cylindrically and spherically aligned, respectively.

For the whole sample stars, the velocity ellipsoid changes from cylindrical alignment at $R\sim4.0$\,kpc to nearly spherical alignment as $R$ increases to $R\sim8.0$\,kpc, and then gradually becomes more cylindrically aligned again toward $R\leq12.0$\,kpc.
The vertical dependence of the alignment also varies across the Galactic disk.
For regions with $R\leq6.0$\,kpc and $R\geq11.0$\,kpc, stars at larger $|Z|$ preferentially show more cylindrical alignment.
In contrast, within the intermediate radial range of $6.0<R<11.0$\,kpc, the velocity ellipsoid becomes progressively closer to spherical alignment with increasing $|Z|$.
In addition, spatial bins with larger velocity dispersions tend to show velocity ellipsoids that are more strongly aligned toward cylindrical configurations.
These results are broadly consistent with recent measurements based on LAMOST-Gaia red clump stars \citep[][]{Sun2023}.

To quantitatively characterize the distribution of the velocity ellipsoid tilt angles, we present the relation between the $\alpha$ and $Z/R$ in Fig.\,3.
Previous studies have shown that the tilt angle distribution can be approximately described by a simple analytical model \citep[e.g.,][]{Binney2014, Budenbender2015, Hagen2019}, given by

\begin{equation}
\label{eq:tiltangle2}
    \alpha = \alpha_{0}\, \mathrm{arctan}(\frac{Z}{R})
\end{equation}
where $\alpha_{0}$ is a fitting parameter.
A value of $\alpha_{0}=1.0$ corresponds to perfect spherical alignment, whereas $\alpha_{0}<1.0$ indicates that the velocity ellipsoid is tilted toward cylindrical alignment.
We note that a perfectly cylindrical alignment ($\alpha_0=0$) is not physically expected, as even a maximally disc-dominated system would be expected to exhibit a non-zero $\alpha_0$.

We use the Python package emcee \citep{Foreman-Mackey2013} to sample the posterior distributions of the model parameters. We adopt a Gaussian likelihood, equivalent to minimizing the weighted least-squares statistic, and employ 100 walkers with 5000 steps each, discarding the first 2000 steps of each chain as burn-in.
All spatial bins containing no fewer than 10 stars are included in the fitting procedure, since these bins still preserve useful information on the velocity ellipsoid alignment, although they generally exhibit relatively larger uncertainties.
In contrast, bins with fewer stars tend to show randomly distributed alignments due to limited statistical significance.
The best-fitting relation is shown as the blue dotted line in the figure, yielding $\alpha_{0}$ = 0.78 $\pm$\,0.05.

Our result indicates that the velocity ellipsoid is preferentially tilted toward cylindrical alignment.
This result is broadly consistent with the local RAVE sample by Binney et al. ({\color{blue}{2014}}), who obtained $\alpha_{0}$\,$\sim$\,0.80.
It is also consistent with the result of $\alpha_{0}$\,=\,0.90\,$\pm$\,0.04 derived by B{\"u}denbender et al. ({\color{blue}{2015}}), as the two measurements agree within the $1\sigma$ uncertainties, with the two error intervals showing substantial overlap.
However, it is slightly larger than that reported for the red clump sample analyzed by Sun et al. ({\color{blue}{2023}}), who reported $\alpha_{0}$\,=\,0.68\,$\pm$\,0.05.
It is slightly smaller than the result of $\alpha_{0}$\,=\,0.95\,$\pm$\,0.01 reported by Everall et al. ({\color{blue}{2019}}).

The study of the alignment of the velocity dispersion tensor can reveal the underlying gravitational potential \citep[e.g.,][]{An2016, Evans2016}.
The results indicate that if the alignment is everywhere near spherical, then the potential must also be spherical (if nonsingular).
If the alignment is everywhere near cylindrical, the potential must be separable in cylindrical coordinates.
In our results with $\alpha_{0}$ = 0.78 $\pm$\,0.05, the alignment is neither spherical nor cylindrical, but in between.
Such coupling between the radial and vertical motions captured in the velocity ellipsoid tilts tells us about the interplay between the disk and halo potentials.
These are in good agreement with recent suggestions \citep[][]{Sun2023}.

To further investigate the metallicity dependence of the tilt of the velocity ellipsoid, we present the $\alpha_{0}$ as a function of stellar [Fe/H] in Fig.\,4.
The result shows a tentative hint of a non-monotonic variation in $\alpha_{0}$ with metallicity.
A simple linear fit to the data yields a slope consistent with zero (0.01 $\pm$ 0.04, p-value $\sim$ 0.97 if weighted fit), indicating no statistically significant linear trend. However, the data points loosely suggest a possible pattern: $\alpha_{0}$ may increase from $\sim$ 0.60 at [Fe/H] = $-$0.90\,dex, peak around $\sim$ 0.70 at [Fe/H] $\sim$ 0.00\,dex, and then slightly decrease towards higher [Fe/H].
Given the large uncertainties in the current sample, we can only report this as a weak, tentative trend rather than a confirmed systematic variation.
Previous studies have reported oscillations in $\alpha_{0}$ with $R$ \citep[e.g.,][]{Sun2023, Sun2024a}, as well as a well-established strong linear relationship between [Fe/H] and $R$ \citep[e.g.,][]{Nordstrom2004, Minchev2018, Sun2024b}.
Thus, the possible variation of $\alpha_{0}$ with [Fe/H] may, at least in part, reflect their respective dependencies on $R$.

To describe the tentative profile of the $\alpha_{0}$--[Fe/H] relation, we adopt a function similar to that used to describe the $\alpha_{0}$--$R$ relation reported by Sun et al. ({\color{blue}{2023}}), as follows:
\begin{equation}
\label{eq:tiltangle2}
    \alpha_{0} = A\,\mathrm{sin}(\omega \mathrm{[Fe/H]} + \Phi) + C
\end{equation}
The best fit plotted as a dotted line in the figure, the result indicates that $\alpha_{0}$ = 0.10\,$\mathrm{sin}$\,(2.44\,[Fe/H]\,+\,1.62)\,+\,0.64.
We emphasize that this sinusoidal fit is phenomenological, intended to capture the apparent non-monotonic shape, and does not necessarily imply a statistically robust physical oscillation in this metallicity range of the whole sample.
In fact, it does not provide a statistically significant improvement over the linear fit (reduced $\chi^2$ decreases only by less than $\sim$ 0.25 for two additional free parameters).

Based on the best fits, we estimate the $\alpha_{0}$ for the solar-metallicity stars ([Fe/H] = 0.0 dex) to be around $\alpha_{0}$ = 0.74 $\pm$ 0.05.
If this tentative trend is real, this is an acceptable result since the spatial distributions of solar-metallicity stars are concentrated in the Galactic disk, those stars are likely significantly affected by the gravitational potential of the baryonic disk of the Galaxy.
That is why the $\alpha_{0}$ is generally smaller than the other results \citep[e.g.,][]{Budenbender2015, Everall2019, Hagen2019}.


\subsection{The Velocity Ellipsoid distributions and their tilts for the Low-[$\alpha$/Fe] and High-[$\alpha$/Fe] populations}

The fact that the metallicity dependence in $\alpha_{0}$ is not only related to the metallicity dependence in $R$, but may also be associated with stellar populations \citep[e.g.,][]{Sun2023, Sun2024a}.
Sun et al. ({\color{blue}{2023}}) reported a clear dependence of $\alpha_{0}$ on stellar populations.
For High-[$\alpha$/Fe] (old) populations, their $\alpha_{0}$ tends to be smaller than that of Low-[$\alpha$/Fe] (young) populations.
These findings suggest that young/Low-[$\alpha$/Fe] stars are generally more consistent with spherical alignment than old/High-[$\alpha$/Fe] stars.
This behavior may reflect the different dynamical evolution of stellar populations with different ages, with older populations potentially retaining stronger signatures of long-term dynamical heating and perturbations.

To further investigate the metallicity dependence of $\alpha_{0}$, we present the velocity ellipsoid distribution of Low-/High-[$\alpha$/Fe] populations in Fig.\,5, while the $\alpha_{0}$--[Fe/H] relations of the two populations are shown in Fig.\,6.

For the Low-[$\alpha$/Fe] stars (see Fig. 5), the velocity ellipsoids show cylindrical alignment in the inner disk region and gradually tend toward spherical alignment in the outer disk region.
For the vertical direction, no significant global trend is observed.
This is the expected result since most of the thin-disk stars are located close to the baryonic disk plane (stellar $|Z|$ $<$ 2.5\,kpc).
The size of the velocity ellipsoid exhibits an overall decreasing trend with increasing $R$ for $R$ $<$ 11 kpc, indicating that both velocity dispersion components gradually decrease with increasing $R$.
However, when $R$ $\geq$ 12.5 kpc, the shape of the velocity ellipsoid becomes rounder, suggesting that $\sigma_{Z}$ tends to flatten or increase slightly with increasing $R$, which may be associated with disk warp behavior \citep[e.g.,][]{Mackereth2019, Sun2024c, Sun2025a, Sun2025b}, although distance uncertainties may also contribute, as the proper-motion-derived vertical velocity dispersion is more sensitive to distance errors than the radial velocity dispersion.

For High-[$\alpha$/Fe] stars, the velocity ellipsoids show spherical alignment at lower Galactic heights and gradually become more cylindrically aligned toward larger Galactic heights, while along the radial direction, no clear global trend is found.
The size of the velocity ellipsoid exhibits a global decreasing trend with $R$, but no obvious trend with $Z$.
In the inner disk region, stars with the largest velocity dispersion generally have velocity ellipsoids preferentially tilted toward cylindrical alignment.
In contrast, for the outer disk region, stars with the largest velocity dispersion generally show more randomly oriented velocity ellipsoids.
This random alignment may be closely related to the complex dynamical heating history of the High-[$\alpha$/Fe] stars, such as merger and/or accretion events \citep[e.g.,][]{Quillen2001, Abadi2003}.
In addition, stars located at larger Galactic heights tend to exhibit velocity ellipsoids that are globally more cylindrically aligned.

We also fit the $\alpha_{0}$ of the two populations with equation (3), and the results are also summarized in Table\,1.
The best-fitting results for the Low-[$\alpha$/Fe] and High-[$\alpha$/Fe] populations are respectively  $\alpha_{0}$ = 0.86 $\pm$ 0.07 and $\alpha_{0}$ = 0.77 $\pm$ 0.07.
These results are in rough agreement with the red clump sample in Sun et al. ({\color{blue}{2023}}), who reported $\alpha_{0}$ = 0.72 $\pm$ 0.08 for thin disk stars and $\alpha_{0}$ = 0.64 $\pm$ 0.06 for thick disk stars.
Our Low-[$\alpha$/Fe] stars are generally more consistent with spherical alignment than High-[$\alpha$/Fe] stars, again suggesting that the influence of the baryonic disk gravitational potential becomes increasingly important for older and more dynamically relaxed stellar populations.

Fig.\,6 displays the $\alpha_{0}$--[Fe/H] relations for Low- and High-[$\alpha$/Fe] populations.
The results indicate that the $\alpha_{0}$ of both Low- and High-[$\alpha$/Fe] populations shows possible oscillatory variations with [Fe/H].
We fit the $\alpha_{0}$--[Fe/H] relations of the two populations using equation (4), with the best-fitting results shown by the red lines in each panel.
The sine-function fits provide a reasonable description of the observed variations, with the Low-[$\alpha$/Fe] and High-[$\alpha$/Fe] populations respectively given by $\alpha_{0}$ = 0.12 sin (3.72 [Fe/H] + 1.09) + 0.75 and $\alpha_{0}$ = 0.20 sin (2.99 [Fe/H] + 1.90) + 0.71.
Based on the best-fit results, we can accurately determine the $\alpha_{0}$ of the solar-like stars ([Fe/H] = 0.0\,dex) for Low-[$\alpha$/Fe] and High-[$\alpha$/Fe] populations are respectively $\alpha_{0}$ = 0.87 $\pm$ 0.07 and $\alpha_{0}$ = 0.91 $\pm$ 0.07.
These values also broadly consistent with the solar neighborhood results reported by previous studies \citep[e.g.,][]{Budenbender2015, Sun2023}.

Sun et al. ({\color{blue}{2023, 2024a}}) reported oscillations in $\alpha_{0}$--$R$ for thin and thick disks, which may be associated with the resonance of the Galactic bar.
For our Low-[$\alpha$/Fe] result, the possible oscillatory variations in $\alpha_{0}$--[Fe/H] may also be related to resonances induced by the Galactic bar, given the strong linear correlation between [Fe/H] and $R$ \citep[e.g.,][]{Nordstrom2004, Minchev2018, Donor2020, Spina2021}.

However, the observed variations in $\alpha_{0}$--[Fe/H] of the High-[$\alpha$/Fe] stars represent a particularly interesting result, because the [Fe/H] of the High-[$\alpha$/Fe] stars does not exhibit a clear trend of variation with $R$ \citep[e.g.,][]{Miranda2016, Sun2024b}.
This suggests that the observed variations in $\alpha_{0}$--[Fe/H] are unlikely to arise solely from the reported variations in $\alpha_{0}$--$R$.
Consequently, the observed variations in $\alpha_{0}$--[Fe/H] of the High-[$\alpha$/Fe] stars may have contributions from other dynamical and evolutionary processes, such as violent heating associated with mergers \citep[e.g.,][]{Quinn1993, Grand2016}, accretion \citep{Abadi2003}, and/or the formation of these stars in an environment shaped by mergers of gas-rich systems and/or a turbulent interstellar medium \citep[ISM; e.g.,][]{Brook2012, Wisnioski2015}.

In addition, $\alpha_{0}$ of the metal-rich ([Fe/H] $>$ 0.2 dex) High-[$\alpha$/Fe] stars show an obvious decreasing trend with $R$, which may indicate star formation associated with the so-called "parallel" model of Galactic chemical evolution model \citep[e.g.,][]{Grisoni2017}, and/or the effects of stellar migration \citep[e.g.,][]{Kubryk2015, Grisoni2017}.

We also find that the significant north-south asymmetry in velocity ellipsoids, consistent with previous studies of vertical asymmetries in the Galactic disk \citep[e.g.,][]{Widrow2012, Everall2019, Sun2023}.
Although the origin of this phenomenon remains unclear, several mechanisms may be related to it, including substructures and stellar streams in the Galactic disk, buckling of the Galactic central bar \citep[e.g.,][]{Saha2013}, disk bending perturbations \citep[e.g.,][]{Gomez2013, Williams2013, Xu2015, Laporte2019}, warp structures \citep[e.g.,][]{Momany2006, Williams2013, Sun2025c}, or possible unrecognized systematic effects in the observational data.

\section{Conclusions}

In this paper, we use over 572,000 stars selected from the GALAH and {\it Gaia} to calculate the metallicity dependence of the tilt of the velocity ellipsoid for different disk populations.
We find that the tilt angles of the velocity ellipsoid are distinct among different stellar populations and vary with stellar population and metallicity.

The tilt angle of different stellar populations are well modelled as $\alpha$ = $\alpha_{0}$\,$\mathrm{arctan}$\,($Z$/$R$), with $\alpha_{0}$ = 0.78 $\pm$ 0.05, $\alpha$ = 0.86 $\pm$ 0.07 and $\alpha$ = 0.77 $\pm$ 0.07, for respectively, the whole sample, Low-[$\alpha$/Fe] population and High-[$\alpha$/Fe] population.
These results indicate that various disk populations are globally tilted towards cylindrical alignment, implying that the deviation from spherical alignment is caused by the gravitational potential of the Galactic baryonic disk, and that the kinematically relaxed stars are mainly dominated by the gravitational potential of the baryonic disk.

Our detailed measurement of the $\alpha_{0}$ -- [Fe/H] relations show variations of $\alpha_{0}$ with [Fe/H] for the different populations.
For the whole sample, this relation can be described as $\alpha_{0}$ = 0.10\,$\mathrm{sin}$\,(2.44\,[Fe/H]\,+\,1.62)\,+\,0.64.
For the Low-[$\alpha$/Fe] and High-[$\alpha$/Fe] populations, the corresponding relations are $\alpha_{0}$ = 0.12\,$\mathrm{sin}$\,(3.72\,[Fe/H]\,+\,1.09)\,+\,0.75 and $\alpha_{0}$ = 0.20\,$\mathrm{sin}$\,(2.99\,[Fe/H]\,+\,1.90)\,+\,0.71, respectively.

The metallicity dependence of $\alpha_{0}$ for the Low-[$\alpha$/Fe] population may be primarily associated with resonances induced by the Galactic bar, while that for the High-[$\alpha$/Fe] population may also have been affected by additional dynamical and evolutionary processes, such as violent heating associated with mergers, accretion, and/or the formation of these stars in an environment shaped by mergers of gas-rich systems and/or a turbulent ISM.

\section*{Acknowledgements}
We thank the anonymous referee for very the useful suggestions to improve the work.
This work is supported by the NSFC project 12603038, 12133002, and the CPSF under grant Nos. GZC20240125, 2024M760240, 2025T180873, 2026M793506, and YJA20250180, and the Tsinghua Shuimu Scholar No. 2025SM415, and Henan Provincial Key Science and Technology Project 262102210236.
H.L. is supported by the National Key R\&D Program of China No. 2023YFB3002502, the National Natural Science Foundation of China under No. 12373006 and 12533004, and the China Manned Space Program with grant No. CMS-CSST-2025-A10.

Guoshoujing Telescope (the Large Sky Area Multi-Object Fiber Spectroscopic Telescope LAMOST) is a National Major Scientific Project built by the Chinese Academy of Sciences. Funding for the project has been provided by the National Development and Reform Commission. LAMOST is operated and managed by the National Astronomical Observatories, Chinese Academy of Sciences.

\appendix

\section{Robustness Tests under Different Uncertainty Treatments}

This appendix presents the comparison of the best-fitting tilt-angle parameters obtained under different uncertainty treatments, including different relative distance uncertainty thresholds, Monte Carlo parallax perturbations based on Gaia uncertainties, and Monte Carlo metallicity perturbations (see Table.\,{\color{blue}{A1}}).

The inferred parameters remain consistent within the statistical uncertainties, demonstrating that our conclusions are insensitive to the adopted uncertainty treatments, including distance selection criteria, parallax uncertainties, and metallicity uncertainties.

\begin{table*}

\caption{Comparison of the best-fitting tilt-angle parameters under different uncertainty treatments.}

\centering
\setlength{\tabcolsep}{3mm}{
\resizebox{2.\columnwidth}{!}{
\begin{tabular}{lllllll}
\hline
\hline
\specialrule{0em}{7pt}{0pt}
Uncertainty treatments  &  $\alpha_{0}$ (total) &   $\alpha_{0}$ (Low-[$\alpha$/Fe])   &  $\alpha_{0}$ (High-[$\alpha$/Fe]) \\[0.07cm]
                                &  (degree)     &    (degree)         &      (degree)     \\
\specialrule{0em}{7pt}{0pt}
\hline
\specialrule{0em}{7pt}{0pt}
$\Delta s$ $<$ 5\%             &  0.78 $\pm$ 0.06  &    0.83 $\pm$ 0.10  &       0.78 $\pm$ 0.07  \\[0.3cm]
$\Delta s$ $<$ 10\%            &  0.78 $\pm$ 0.05  &    0.86 $\pm$ 0.07  &       0.77 $\pm$ 0.07  \\[0.3cm]
$\Delta s$ $<$ 15\%            &  0.76 $\pm$ 0.06 &    0.89 $\pm$ 0.07  &       0.76 $\pm$ 0.06  \\[0.3cm]
$\Delta s$ $<$ 10\% $+$ Monte Carlo parallax perturbations using Gaia uncertainties  &  0.77 $\pm$ 0.06      &    0.85 $\pm$ 0.08 &       0.76 $\pm$ 0.07  \\[0.3cm]
$\Delta s$ $<$ 10\% $+$ Monte Carlo [Fe/H] uncertainty    &  0.79 $\pm$ 0.05 &    0.86 $\pm$ 0.07  &       0.78 $\pm$ 0.07  \\
\specialrule{0em}{7pt}{0pt}
\hline 
\specialrule{0em}{7pt}{0pt}
\end{tabular}}
}
\label{tab:datasets}
\end{table*}


\begin{thebibliography}{}

\bibitem[Abadi et al.(2003)]{Abadi2003} Abadi, M. G., Navarro, J. F., Steinmetz, M., Eke, V. R.\ 2003, \apj, 597, 21

\bibitem[Ablimit et al.(2020)]{Ablimit2020} Ablimit, I., Zhao, G., Flynn, C., Bird, S. A.\ 2020, \apj, 895, 12

\bibitem[An \& Evans(2016)]{An2016} An, J. \& Evans, N. W.\ 2016, \apj, 816, 35

\bibitem[Bensby, Feltzing \& Oey(2014)]{Bensby2014} Bensby, T., Feltzing, S., \& Oey, M. S.\ 2014, \aap, 562, A71

\bibitem[Binney \& Merrifield(1998)]{Binney1998} Binney, J., \& Merrifield, M.\ 1998, Galactic Astronomy (Princeton, NJ: Princeton Univ. Press)

\bibitem[Binney \& Tremaine(2008)]{Binney2008} Binney, J., \& Tremaine, S.\ 2008, Galactic Dynamics (2nd ed.; Princeton, NJ: Princeton Univ. Press)

\bibitem[Binney et al.(2014)]{Binney2014} Binney, J., Burnett, B., Kordopatis, G., et al.\ 2014, \mnras, 439, 1231

\bibitem[Bland-Hawthorn \& Gerhard(2016)]{Bland-Hawthorn2016} Bland-Hawthorn, J., Gerhard, O.\ 2016, \araa, 54, 529

\bibitem[Bovy et al.(2016)]{Bovy2016} Bovy, J., Rix, H. W., Schlafly, E. F., et al.\ 2016, \apj, 823, 30

\bibitem[Brook et al.(2012)]{Brook2012} Brook, C. B., Stinson, G. S., Gibson, B. K., et al.\ 2012, \mnras, 426, 690

\bibitem[B{\"u}denbender et al.(2015)]{Budenbender2015} B{\"u}denbender, A., van de Ven, G., Watkins, L. L.\ 2015, \mnras, 452, 956

\bibitem[Buder et al.(2021)]{Buder2021} Buder, S., Sharma, S., Kos, J., et al.\ 2021, \mnras, 506, 1, 150. doi:10.1093/mnras/stab1242

\bibitem[De Silva et al.(2015)]{De Silva2015} De Silva, G.~M., Freeman, K.~C., Bland-Hawthorn, J., et al.\ 2015, \mnras, 449, 3, 2604

\bibitem[Donor et al.(2020)]{Donor2020} Donor, J.,  Frinchaboy, P. M., Cunha, K., et al.\ 2020, \aj, 159, 199

\bibitem[Eddington(1915)]{Eddington1915} Eddington, A. S.\ 1915, \mnras, 76, 37

\bibitem[Eilers et al.(2019)]{Eilers2019} Eilers, A. C., Hogg, D. W., Rix, H. W., et al.\ 2019, \apj, 871, 120

\bibitem[Evans et al.(2016)]{Evans2016} Evans, N. W., Sanders, J. L., Williams, A. A., et al.\ 2016, \mnras, 456, 4506

\bibitem[Everall et al.(2019)]{Everall2019} Everall, A., Evans, N. W., Belokurov, V., and Sch{\"o}nrich, R.\ 2019, \mnras, 489, 910

\bibitem[Foreman-Mackey et al.(2013)]{Foreman-Mackey2013} Foreman-Mackey, D., Hogg, D. W., Lang, D., et al.\ 2013, PASP, 125, 306


\bibitem[Gaia Collaboration et al.(2022a)]{Gaia2022a} Gaia Collaboration, Creevey, O. L.,  Sarro, L. M.,  Lobel, A., et al.\ 2022, arXiv:2206.05870 

\bibitem[Gaia Collaboration et al.(2022b)]{Gaia2022b} Gaia Collaboration, Recio-Blanco, A., Kordopatis, G., et al.\ 2022, arXiv:2206.05534

\bibitem[G{\' o}mez et al.(2013)]{Gomez2013} G{\' o}mez, F. A., Minchev, I., O’Shea, B. W., Beers, T. C., Bullock, J. S., Purcell, C. W.\ 2013, \mnras, 429, 159

\bibitem[Grand et al.(2016)]{Grand2016} Grand, R. J. J., Springel, V., G{\'o}mez, F. A., et al.\ 2016, \mnras, 459, 199

\bibitem[Grisoni et al.(2017)]{Grisoni2017} Grisoni, V., Spitoni, E., Matteucci, F., et al.,\ 2017, \mnras, 472, 3637

\bibitem[Guiglion et al.(2015)]{Guiglion2015} Guiglion, G., Recio-Blanco, A., de Laverny, P., et al.\ 2015, \aap, 583, A91

\bibitem[Hagen et al.(2019)]{Hagen2019} Hagen, J. H. J., Helmi, A., de Zeeuw, P. T.,  and Posti, L.\ 2019, \aap, 629, A70

\bibitem[Hayden et al.(2020)]{Hayden2020} Hayden, M. R., Bland-Hawthorn, J., Sharma, S., et al.\ 2020, \mnras, 493, 2952

\bibitem[Haywood et al.(2013)]{Haywood2013} Haywood, M., Di Matteo, P., Lehnert, M. D., et al.\ 2013, \aap, 560, A109

\bibitem[Huang et al.(2015)]{Huang2015} Huang, Y., Liu, X. W., Yuan, H. B., et al.\ 2015, \mnras, 449, 162

\bibitem[Huang et al.(2016)]{Huang2016} Huang, Y., Liu, X. W., Yuan, H. B., et al.\ 2016, \mnras, 463, 2623

\bibitem[Kubryk, Prantzos \& Athanassoula(2015)]{Kubryk2015} Kubryk, M., Prantzos, N., Athanassoula, E.,\ 2015, \aap, 580, A126

\bibitem[Laporte et al.(2019)]{Laporte2019} Laporte, C. F. P., Minchev, I., Johnston, K. V., G{\' o}mez, F. A.\ 2019, \mnras, 485, 3134



\bibitem[Lee et al.(2011)]{Lee2011} Lee, Y. S., Beers, T. C., An, D., et al.\ 2011, \apj, 738, 187


\bibitem[Mackereth et al.(2019)]{Mackereth2019} Mackereth, J. T., Bovy, J., Leung, H. W., et al.\ 2019, \mnras, 489, 176



\bibitem[Minchev et al.(2015)]{Minchev2015} Minchev, I., Martig, M., Streich, D., et al.\ 2015, \apj, 804, L9


\bibitem[Minchev et al.(2018)]{Minchev2018} Minchev, I., Anders, F., Recio-Blanco, A., et al.\ 2018, \mnras, 481, 2, 1645


\bibitem[Miranda et al.(2016)]{Miranda2016} Miranda, M.~S., Pilkington, K., Gibson, B.~K., et al.\ 2016, \aap, 587, A10

\bibitem[Momany et al.(2006)]{Momany2006} Momany, Y., Zaggia, S., Gilmore, G., et al.\ 2006, \aap, 451, 515

\bibitem[Nordstrom et al.(2004)]{Nordstrom2004} Nordstr{\"o}m, B., Mayor, M., Andersen, J., et al.\ 2004, \aap, 418, 989

\bibitem[Quillen \& Garnett(2001)]{Quillen2001} Quillen, A. C., \& Garnett, D. R. 2001, in ASP Conf. Ser. 230, Galaxy Disks and Disk Galaxies, ed. J. G. Funes \& E. M. Corsini (San Francisco, CA: ASP), 87

\bibitem[Quinn et al.(1993)]{Quinn1993} Quinn, P. J., Hernquist, L., \& Fullagar, D. P.\ 1993, \apj, 403, 74


\bibitem[Recio-Blanco et al.(2014)]{Recio-Blanco2014} Recio-Blanco, A., de Laverny, P., Kordopatis, G., et al.\ 2014, \aap, 567, A5

\bibitem[Recio-Blanco et al.(2022)]{Recio2022} Recio-Blanco, A., de Laverny, P., Palicio, P.~A., et al.\ 2022, arXiv:2206.05541.

\bibitem[Reid \& Brunthaler(2004)]{Reid2004} Reid, M. J., \& Brunthaler, A.\ 2004, \apj, 616, 872

\bibitem[Reid et al.(2014)]{Reid2014} Reid, M. J., Menten, K. M., Brunthaler, A., et al.\ 2014, \apj, 783, 130

\bibitem[Saha, Pfenniger \& Taam(2013)]{Saha2013} Saha, K., Pfenniger, D., Taam, R. E.\ 2013, \apj, 764, 123

\bibitem[Sch{\"o}nrich et al.(2010)]{Schonrich2010} Sch{\"o}nrich, R., Binney, J., \& Dehnen, W.\ 2010, \mnras, 403, 1829

\bibitem[Sch{\"o}nrich(2012)]{Schonrich2012} Sch{\"o}nrich, R.\ 2012, \mnras, 427, 274

\bibitem[Sch{\"o}nrich \& Dehnen(2018)]{Schonrich2018} Sch{\"o}nrich, R., \& Dehnen, W.\ 2018, \mnras, 478, 3809

\bibitem[Smith et al.(2009)]{Smith2009} Smith, M. C., Evans, N. W., An, J. H.\ 2009, \apj, 698, 1110

\bibitem[Spina et al.(2021)]{Spina2021} Spina, L., Ting, Y. S., De Silva, G. M., et al.\ 2021, \mnras, 503, 3279


\bibitem[Sun et al.(2020)]{Sun2020} Sun, W. X., Huang, Y., Wang, H. F., et al.\ 2020, \apj, 903, 12

\bibitem[Sun et al.(2023)]{Sun2023} Sun, W. X., Shen, H., Liu, X. W.\ 2023, \apj, 952, 163

\bibitem[Sun et al.(2024a)]{Sun2024a} Sun, W. X., Shen, H., Jiang, B. W., \& Liu, X. W. .\ 2024a, \apjl, 965, L9 

\bibitem[Sun et al.(2024b)]{Sun2024b} Sun, W. X., Shen, H., Jiang, B. W., \& Liu, X. W. .\ 2024b, \apjs, 272, 8

\bibitem[Sun et al.(2024c)]{Sun2024c} Sun, W. X., Huang, Y., Shen, H., et al.\ 2024c, \apj, 961, 141

\bibitem[Sun et al.(2025a)]{Sun2025a} Sun, W. X., Shen, H., Jiang, B. W., \& Liu, X. W. .\ 2025a, \apj, 979, 103

\bibitem[Sun et al.(2025b)]{Sun2025b} Sun, W. X., Shen, H., Jiang, B. W., \& Liu, X. W. .\ 2025b, \apj, 985, 232

\bibitem[Sun et al.(2025c)]{Sun2025c} Sun, W. X., Shen, H., Jiang, B. W., \& Liu, X. W. .\ 2025c, \apjl. 978, L24


\bibitem[Widrow et al.(2012)]{Widrow2012} Widrow, L.~M., Gardner, S., Yanny, B., et al.\ 2012, \apjl, 750, 2, L41.

\bibitem[Williams et al.(2013)]{Williams2013} Williams, M. E. K., Williams, M., Binney, J., et al.\ 2013, \mnras, 436, 101

\bibitem[Wisnioski et al.(2015)]{Wisnioski2015} Wisnioski, E., F{\"o}rster Schreiber, N. M., Wuyts, S., et al.\ 2015, \apj, 799, 209



\bibitem[Xu et al.(2015)]{Xu2015} Xu, Y., Newberg, H. J., Carlin, J. L., et al.\ 2015, \apj, 801, 105


\end{thebibliography}
\end{document}